\documentclass{emulateapj}
\usepackage{apjfonts}

\shorttitle{HD 32297 at multiple wavelengths} 
\shortauthors{Debes et al.}
\begin{document}
\title{Interstellar Medium Sculpting of the HD~32297 Debris Disk}
\author{John H. Debes\altaffilmark{1,2},Alycia J. Weinberger\altaffilmark{3}, Marc J. Kuchner\altaffilmark{1}}

\altaffiltext{1}{Code 667, Goddard Space Flight Center, Greenbelt, MD 20771}
\altaffiltext{2}{NASA Postdoctoral Program Fellow}
\altaffiltext{3}{Department of Terrestrial Magnetism, Carnegie Institution of Washington, Washington, DC 20015}

\begin{abstract}
We detect the HD~32297 debris disk in scattered light at 1.6 and 2.05 \micron. We use these new observations together with a previous scattered light image of the disk at 1.1~\micron\ to examine the structure and scattering efficiency of the disk as a function of wavelength.  In addition to surface brightness asymmetries and a warped morphology beyond $\sim$1\farcs5 for one lobe of the disk, we find that there exists an asymmetry in the spectral features of the grains between the northeastern and southwestern lobes.  The mostly neutral color of the disk lobes imply roughly 1~\micron-sized grains are responsible for the scattering.    We find that the asymmetries in color and morphology can plausibly be explained by HD~32297's motion into a dense ISM cloud at a relative velocity of 15~km s$^{-1}$.  We model the interaction of dust grains with HI gas in the cloud.  We argue that supersonic ballistic drag can explain the morphology of the debris disks of HD~32297, HD~15115, and HD~61005.
\end{abstract}

\keywords{stars:individual(HD 32297)---stars:individual(HD 15115)---stars:individual(HD 61005)---circumstellar disks---methods:n-body simulations---ISM}

\section{Introduction}
Debris disks are created by the outgassing and collisions of planetesimals that may resemble analogues to the comets, asteroids, and dwarf planets in our Solar System.  Debris disks show us the conditions and environments of planet building for stars different than the Sun through their structure and mineralogy.  

As more disks are resolved at multiple wavelengths, they reveal their intrinsic complexity.  Both the morphology and spectra of disks change in ways that are not fully understood.  In order to probe the intrinsic properties of a disk, effects that are due to the surrounding environment of a star must be accounted for.

HD~32297, an A5 star at 112$\pm$11~pc with a recently discovered edge-on debris disk, appears to be a promising object of study.  First noted as a star with an IRAS excess \citep{silverstone00}, \citet[][Hereafter S05]{schneider05} discovered an edge-on disk structure in scattered light at 1.1\micron.  The disk extended out to $\sim$400~AU and showed surface brightness asymmetries between the two lobes of the disk.  This morphology appears to be common among known edge-on disks like $\beta$-Pictoris and HD~15115, which both show similar asymmetries \citep{kalas95,golimowski06,kalas07,debes08}.  The total L$_{IR}$/L$_\star$=0.27\% makes the disk optically thin S05.

 Observations of HD~32297's outer disk in the R band revealed that it may be interacting with the surrounding ISM.  The disk appeared significantly warped at larger separations from the star \citep[][hereafter K05]{kalas05}.  The NE half of the disk appeared to bow away from the midplane of the disk.  An extrapolated surface brightness profile implied blue colors for the disk which K05 interpreted as resulting from sub-micron silicate dust grains.  K05 suggested that the bowed structure results from an ISM-disk interaction.

The disk was also spatially resolved at 12 and 18\micron\ in ground based observations with Gemini \citep{moerchen07,fitzgerald07}.  There is a suggestion that at distances $>$0\farcs75 the disk is brighter on the NE side, i.e. reversed from the asymmetry seen in scattered light images \citep{moerchen07}.  Models of the thermal emission and brightness temperature curves as a function of distance based on the 12\micron/18\micron\ colors suggest that the disk has a ring-like structure with an interior clearing out to $\sim$60-80~AU \citep{fitzgerald07,moerchen07}. 

Finally, HD~32297's disk has been resolved with CARMA at 1.3~mm.  The CARMA image shows a marginally resolved asymmetry in the disk that appears to match asymmetries seen in scattered light \citep{maness08}.  Furthermore, the emission at 1.3~mm is well in excess of that inferred by shorter wavelength thermal emission.  Models of the disk so far rely on analytic models of dust grain optical behavior representative of ``dirty ice'' mixtures, but there are few constraints on the true composition of the dust grains in the disk; the presumption of water ice is premature \citep{fitzgerald07,maness08}.

In addition to dust, a small amount of gas is present in the HD~32297 system.  Based on the detection of narrow circumstellar Na~I absorption in the spectrum of HD~32297, \citet{redfield07} determined that at most 0.3M$_{\oplus}$ of gas resides within the disk. The CARMA observations also provided an upper limit of 95 M$_\oplus$ of gas \citep{maness08}.  

In this article we present two new NICMOS coronagraphic scattered light images of the disk  at 1.6 and 2.05\micron.  We examine how the morphology of the disk changes with wavelength and how that might lead to new constraints on the composition of the dust in the disk.  The properties of the disk suggest that it is interacting with the surrounding ISM.   We model this interaction and explore how it affects the observed properties of the disk.  In \S \ref{s:obs} we present the observations of HD~32297, in \S \ref{s:analysis} we measure the morphology, surface brightness, and scattering efficiency of the dust as a function of wavelength.  Finally, we present models of the interaction between the ISM and HD~32297, HD~15115, and HD~61005 in \S \ref{s:modeling}.
 
\section{Observations}
\label{s:obs}
We observed
HD~32297 with the Hubble Space Telescope's (HST) Near-Infrared
and Multi-Object Spectrometer (NICMOS) \citep{thompson98} 
using the coronagraph in Camera 2.  We
used the F160W and F205W filters with central wavelengths of 1.6 and 2.05\micron.  Additionally, we re-analyzed archival data of HD~32297 in the F110W filter taken by S05  All new observations were taken on 3 October 2007
for HD~32297.

In between visits to HD~32297 we observed HD~31489, the point-spread function (PSF) reference.  Accurate disk photometry requires a well constrained scaling ratio between the target star and its PSF reference.  To achieve the most accurate scaling, over the three visits we obtained three short 0.6s exposures of both the target and PSF reference star without the coronagraph in both filters.  

In F160W, the reference saturated in each 0.6s exposure.  To determine the proper scaling factor for F160W, we used an annulus centered on each star with a radius between 6 to 21 pixels to determine the total counts/s measured.  In F205W, this was not an issue for either star, and so we used a simple circular aperture with a width of 21 pixels.  In F160W we found the ratio of flux densities for HD~32297 to HD~31489 to be 0.54$\pm$0.02.  In F205W, we found a ratio of 0.53$\pm$0.01.

We also used the archival observations of HD~32297 and HD~83870.  HD~83870 was reported in S05 to yield the best PSF subtraction using pixel offsets of [+0.304,+0.143] and a ratio of HD~32297 to HD~83870 of 0.440$\pm$0.005.  Short exposures of both stars were also available, though both were saturated at their cores.  We re-determined the scaling using the same procedure as we used for the F160W filter and obtained a ratio of 0.441$\pm$0.005, in excellent agreement with that determined by S05.

In order to determine the best subtraction we minimized a chi-squared metric on a 
region of the target image dominated by the star's diffraction spikes.  We assumed that good
subtraction of the diffraction spikes corresponded to the best subtraction
of the PSF within the region of interest \citep{cnc03}.  We iteratively created subtractions for combinations of scaling and pixel offsets until we found an image that produced the lowest chi-squared measure.  We searched within 1-$\sigma$ of the scaling ratios we determined above and within $\pm$1~pixel to find the best x and y pixel offsets. 

To quantify the systematic effects on the photometry, we repeated the subtractions varying the PSF scalings and offsets by $\pm$1 $\sigma$ from the minimum chi-square solution found above. Using a rectangular photometric aperture matched to the size of the disk, we found the standard deviation in the disk flux densities from this suite of subtractions.  We then propagated this uncertainty into the total uncertainty in the flux density of the disk per pixel.

\begin{figure}
\plotone{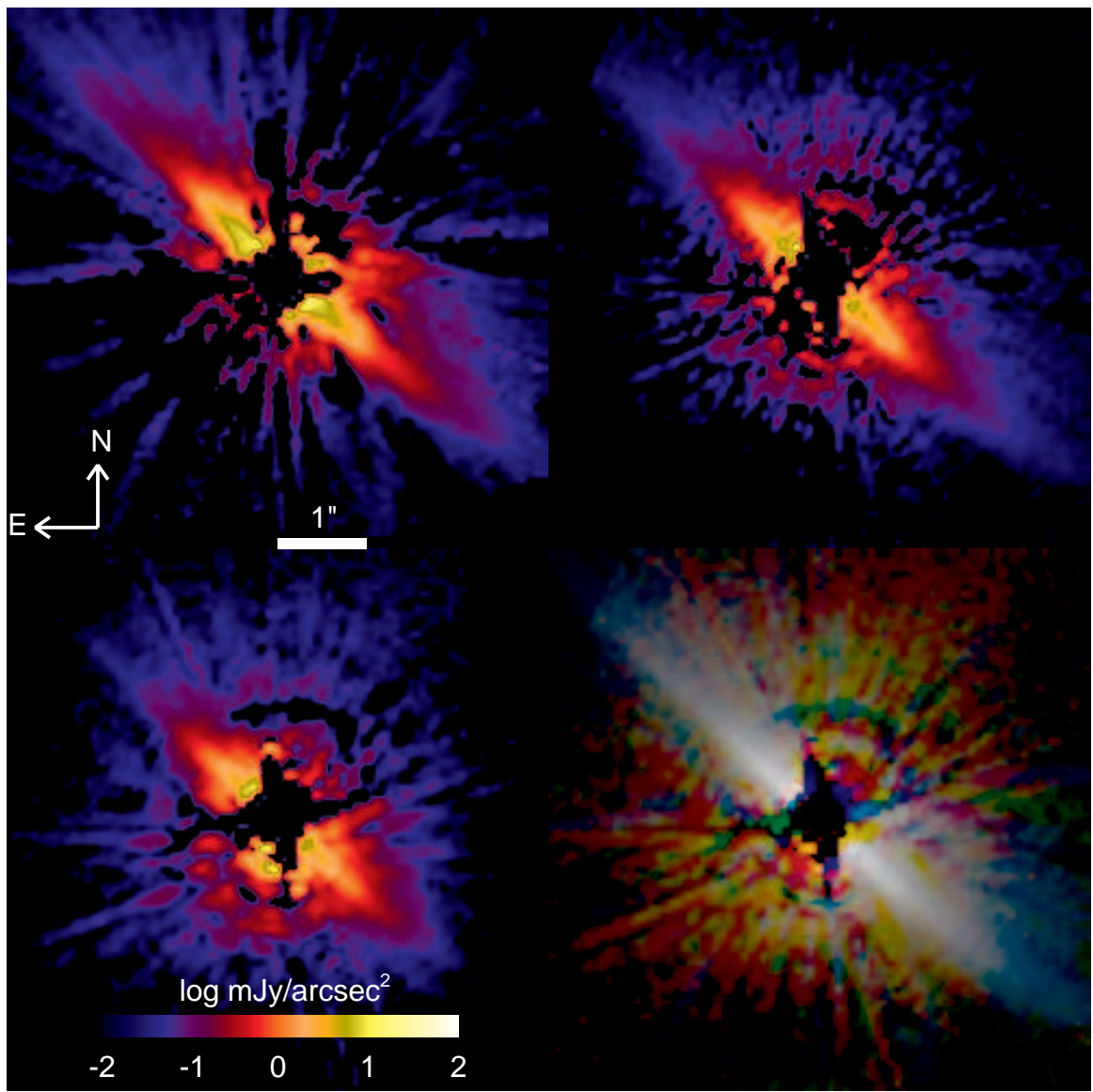}
\caption{\label{fig:ims} Logarithmically scaled, final combined NICMOS coronagraphic images of the HD~32297 debris disk at F110W (1.1\micron, upper left), F160W (1.6\micron, upper right), and F205W (2.05\micron, lower left).  In the lower right corner of the figure is a logarithmically scaled color combined image of the disk, where blue is F110W, green is F160W, and red is F205W.}
\end{figure}

We observed HD~32297 at two telescope orientations differing by
24$^\circ$.  This is essentially an azimuthal dither that allows astronomical
objects that rotate with the field of view to be distinguished from rotationally invariant
instrumental artifacts.    The two
differentially-oriented PSF-subtracted images in each filter band were geometrically rectified and corrected for the linear optical distortion (X$_{scale}$= 75.95 mas pixel$^{-1}$, Y$_{scale}$=75.42 mas pixel$^{-1}$) at the Camera 2 coronagraphic focus.  The geometrically corrected image pairs were rotated about the location of the occulted target (as determined through the target acquisition process) and averaged together.  The final, combined images are shown in Figure \ref{fig:ims}.  

In practice, instrument artifacts are rarely constant in time, so subtraction residuals remain and can contaminate photometric measurements of a disk.  We therefore checked that the total disk flux measurement at each telescope orientation was consistent to the above quoted uncertainties.

\section{Results}
\label{s:analysis}

\subsection{Disk Geometry}
To measure any warping of the HD32297 disk system, we calculated the disk position angle (PA) as a function of radial distance.  To calculate the PA, we performed the following analysis.  First, we separated the images at each spacecraft orientation for all three filters to create six independent images of the disk.  In each image, we highlighted the midplane of the disk by performing an unsharp mask; we subtracted a Gaussian-smoothed disk image from the original image.  Then we used a rectangular aperture of width 3 pixels in the radial direction and height 8 pixels in the direction perpendicular to the midplane of the disk over the disk image.  We centered the aperture on the disk midplane at radial distances of 6 to 40 pixels.  In each aperture we summed the surface brightness  along the 3 pixels in the radial direction radial direction and found the center of the disk by fitting the resulting cross-section with a Gaussian.  For each radial distance we took a median of the six independent measures of the PA from each image and calculated an uncertainty based on the standard deviation of the measures.  The NE lobe has a median PA of 46.3$\pm$1.3 (i.e., 226$^\circ\pm1^\circ$ and the SW lobe has a median PA of 229$^\circ\pm2^\circ$.  Our PA measurement differs from that reported in S05.  However, the F110W PA was corrected to 47.6$^\circ\pm1$ (i.e., 227.6$^\circ\pm1$) in K05 by private communication from G. Schneider.   

\begin{figure}
\plotone{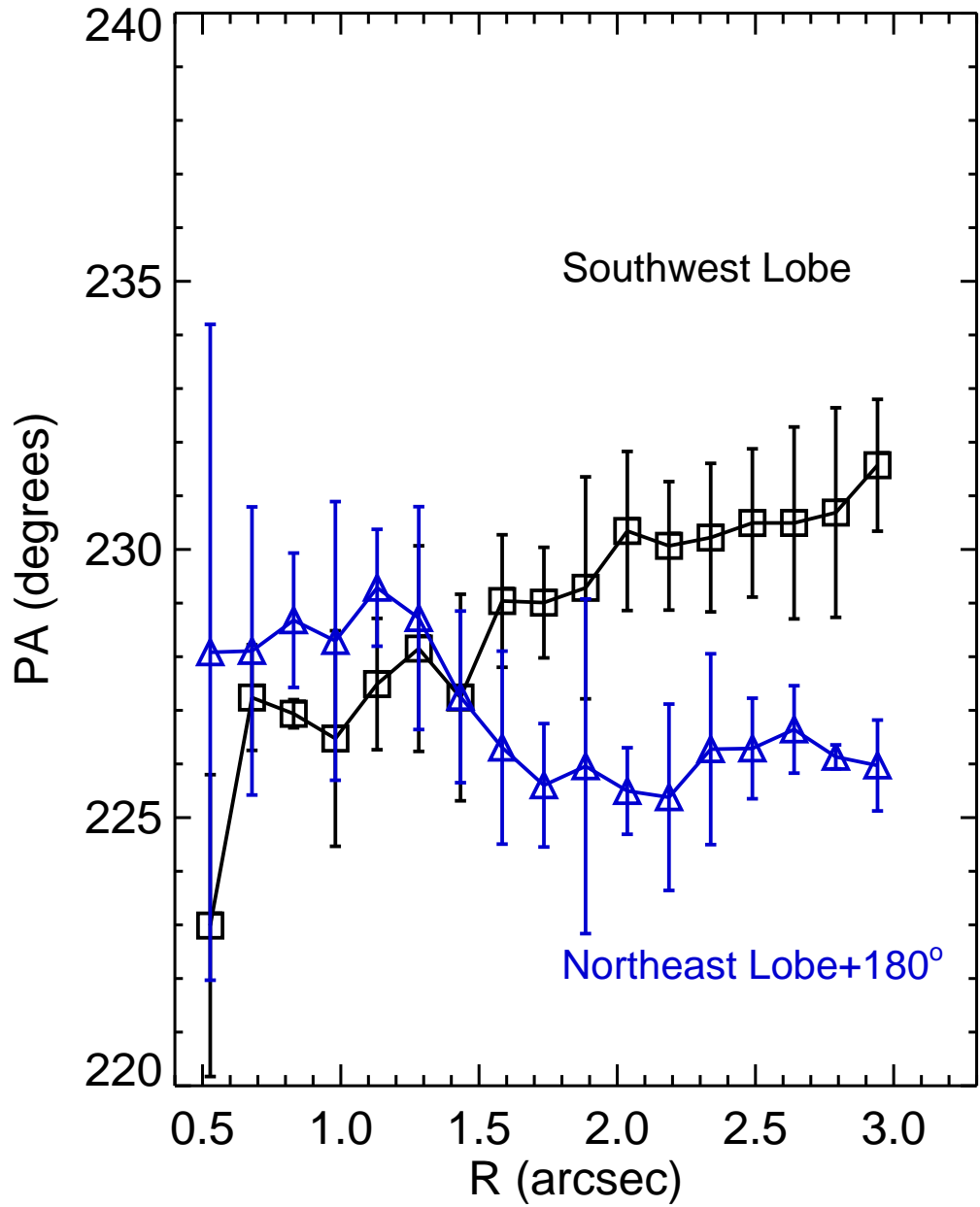}
\caption{\label{fig:pa} Median measure of the position angle (PA) of each lobe of the HD~32297 debris disk as a function of distance from the star.  A significant linear trend in the Southwest lobe is seen.}
\end{figure}

The PA as a function of radial distance as measured from the stellar center is shown in Figure \ref{fig:pa}.  The SW lobe of the disk shows a significant non-zero slope to the PA as a function of distance from the star, i.e., the disk midplane is curved.  The NE lobe shows that the PA is uniform to within the uncertainties.  However, visual inspection of the outer edge isophotes of the NE lobe with respect to the average PA measurements suggests that there might be some slight warping.  

To determine the extent to the warping of the SW lobe, we performed a least squares linear fit to the PA as a function of radial distance from the star.  We found a slope of 2.1$\pm$0.4$^\circ$/\arcsec.  K05 found significantly different PAs for both lobes compared to S05; they reported a PA of 245$^\circ\pm2^\circ$ for the SW lobe but they did not specify at what distance this was measured.  However, if we extrapolate the curvature we see out to 8\arcsec, the radial midpoint of where the disk was detected in R, we would expect the PA to be 246.1$^\circ\pm$3.4$^\circ$, consistent with what was reported in K05.

\subsection{Surface Brightness Profiles}
\label{sec:sb}
To study the surface brightness profiles of the HD~32297 disk we placed 3$\times$3 pixel apertures on the images centered on the disk midplane.  We  measured the surface brightness of the disk in each aperture as a function of distance from the star and wavelength for both disk lobes.  We used background apertures placed just beyond the disk to estimate the level of any systematic uncertainties.  These background apertures were located at similar distances from the star as the photometric apertures.

\begin{figure}
\plotone{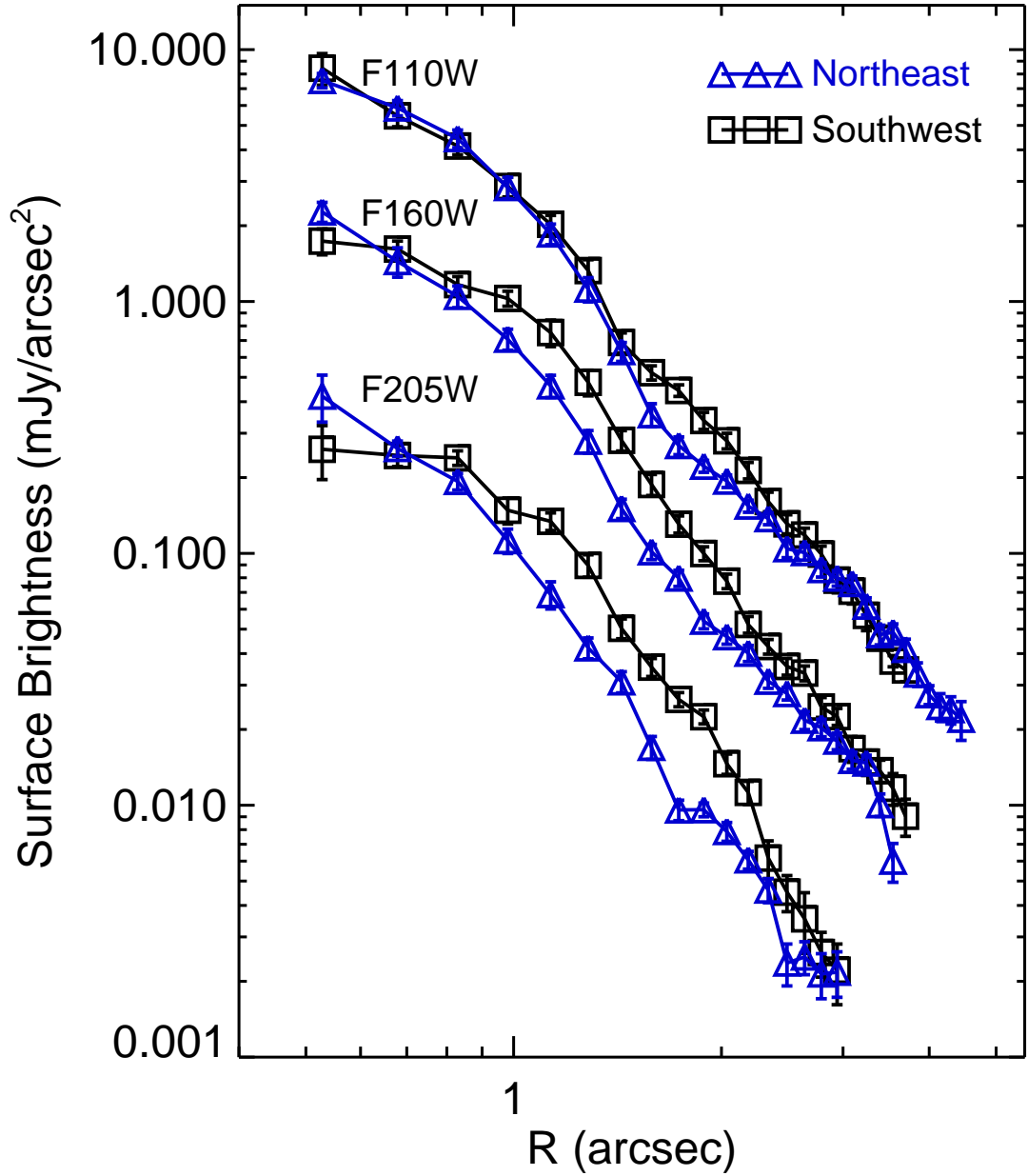}
\caption{\label{fig:sb1} Surface brightness profiles of the Northeast and Southwest lobes of the HD~32297 debris disk in the F110W, F160W, and F205W filters.  The F160W filters are scaled by a factor of 0.5, while the F205W filters are scaled by a factor of 0.125 for clarity.  The surface brightnesses were measured in 3$\times$3 pixel apertures.  A surface brightness asymmetry exists between roughly 1\farcs5 to 3\arcsec in all three filters, with the asymmetry becoming more pronounced at longer wavelengths.}
\end{figure}

Figure \ref{fig:sb1} shows the profiles in each filter.  Overall the profiles have a qualitatively similar shape.  Of particular interest is the amount of asymmetry between the lobe surface brightnesses; S05 noted that the NE lobe of the disk was dimmer than the SW lobe beyond 1\farcs7.  This asymmetry is confirmed at new wavelengths by our data.   We find that the asymmetry is more prominent at F160W and F205W, extending from about 1\arcsec to 2\farcs5.

The surface brightness profiles in both lobes and at all wavelengths cannot be described by a single power-law, rather they are best fit with broken power-laws.  We tabulate the best fitting parameters for the power-laws in Table \ref{tab:pl} along with the radius of the break (R$_{\mathrm{break}}$), the power-law index before R$_{\mathrm{break}}$ $\alpha$ and the power-law index exterior to R$_{\mathrm{break}}$.  The radius at which the break occurs is slightly different between the two lobes, occuring closer to 90~AU on the NE lobe and 110~AU on the SE lobe of the disk, though this is at the limit of our spatial resolution and may not be significant.  Both lobes are characterized by a slope less than r$^{-2}$ in the interior of the disk indicating a rise in the optical depth of scatterers, followed by a drop-off in surface brightness of $\sim$r$^{-3}$ beyond the break corresponding to a slow drop-off in optical depth beyond the break.  

\subsection{Scattering Efficiency vs. Wavelength}
The next step in our analysis is to infer some scattering properties for the dust grains in orbit around HD~32297.  While resolved thermal images of the disk exist, it is not clear if any of the observations to date constrain the composition of the material in the disk \citep{silverstone00, moerchen07, fitzgerald07,maness08}.  The SED at 12\micron\ shows no indication of emission around the strong 10\micron\ silicate line, typical of colder debris disks, or debris disks with larger particles \citep{moerchen07,fitzgerald07}.  In HD~32297's case, the temperature of the dust is estimated to be $\sim$170-180~K, hot enough to produce a silicate line if small silicate grains are present \citep{moerchen07,fitzgerald07}.  For these types of debris disks in particular, scattered light images provide one of the few avenues with which to infer dust composition.

Translating observed surface brightness profiles into information about the dust grains present requires removing the spectral contribution from the star itself to reveal the spectral character of the disk.  In practical terms, the scattering efficiency of dust grains, $Q_{sca}$, is proportional to the ratio of the surface brightness of the disk (SB$_\nu$) to the flux density of the central star (F$_\nu$).  We will measure $Q_{sca}$ by dividing our images (SB$_\nu$) by the F$_\nu$.

For HD~32297, the direct imaging in each filter allows us to accurately estimate the flux density of HD~32297 for the different wavelengths observed.  Since HD~32297 was not saturated in the F160W and F205W filters for our 0.6s exposures, we found that F$_\nu$=0.93$\pm$0.02~Jy and 0.61$\pm$0.01~Jy for F160W and F205W respectively. 

The F110W images pose a slight problem for estimating F$_\nu$ since HD~32297 saturated the detector for the 0.6s exposures.  However, we can use the wings of the PSF to infer the total flux density of the star.  We modeled the HD~32297 PSF with the tinytim software package (V6.2)\footnote{http://www.stsci.edu/software/tinytim/tinytim.html}, assuming an A5 star.  We used the same annulus described in \S \label{sec:obs} for determining our PSF scalings to calculate an aperture correction in our synthetic images between the flux in the wings to the total flux in the PSF.  We assumed a 5\% error in the multiplicative correction term, or 9.6$\pm$0.5.  Our resulting flux density in F110W is 1.55$\pm$0.08~Jy.  S05 estimated the F110W flux density of HD~32297 using SYNPHOT.  They estimated a flux density of 1.46$\pm$0.04~Jy, consistent with our measurement uncertainties.  

\begin{figure}
\plotone{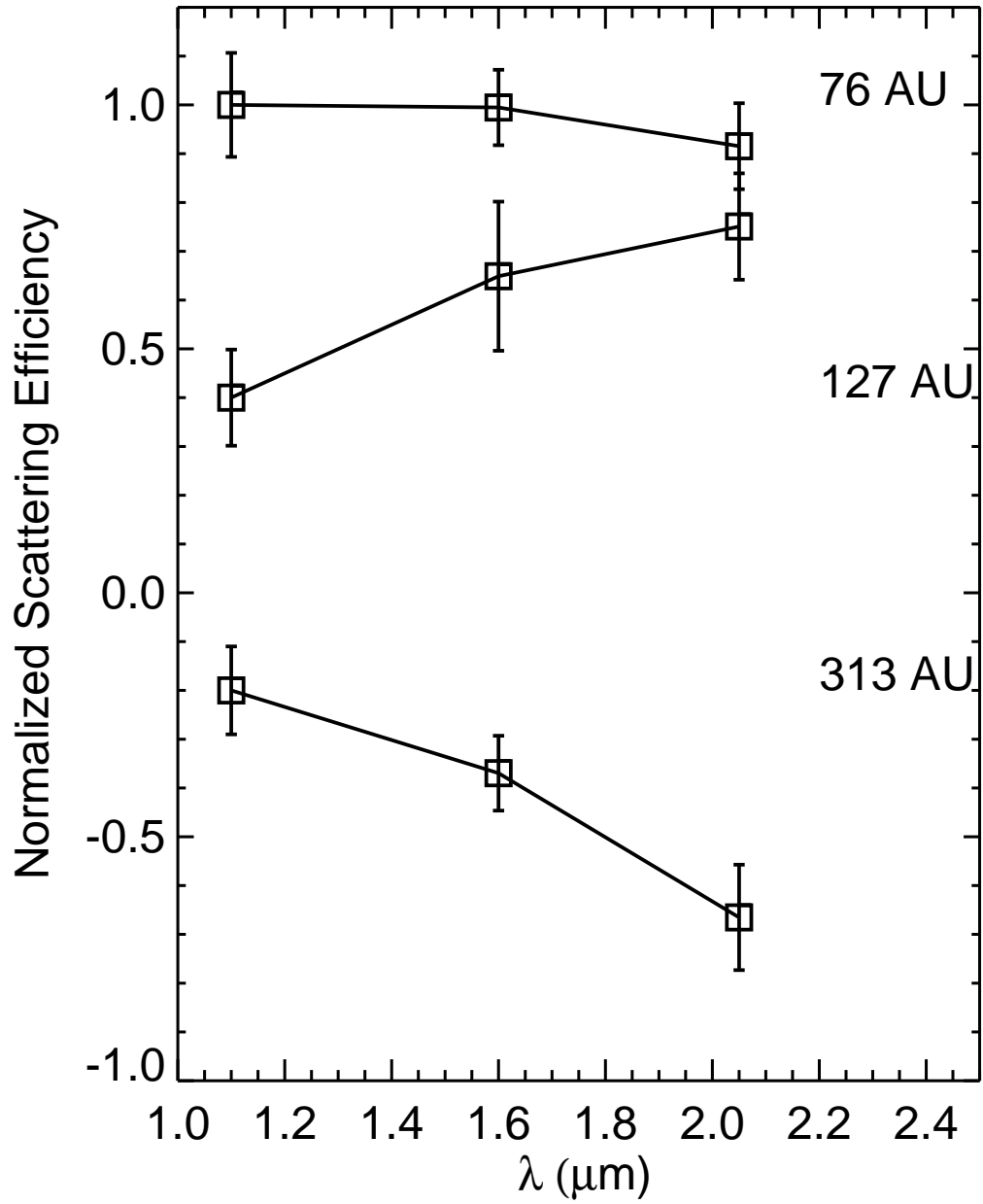}
\caption{\label{fig:color1} Scattering efficiencies at 3 different distances in the HD~32297 disk for the Northeastern lobe.}
\end{figure}

\begin{figure}
\plotone{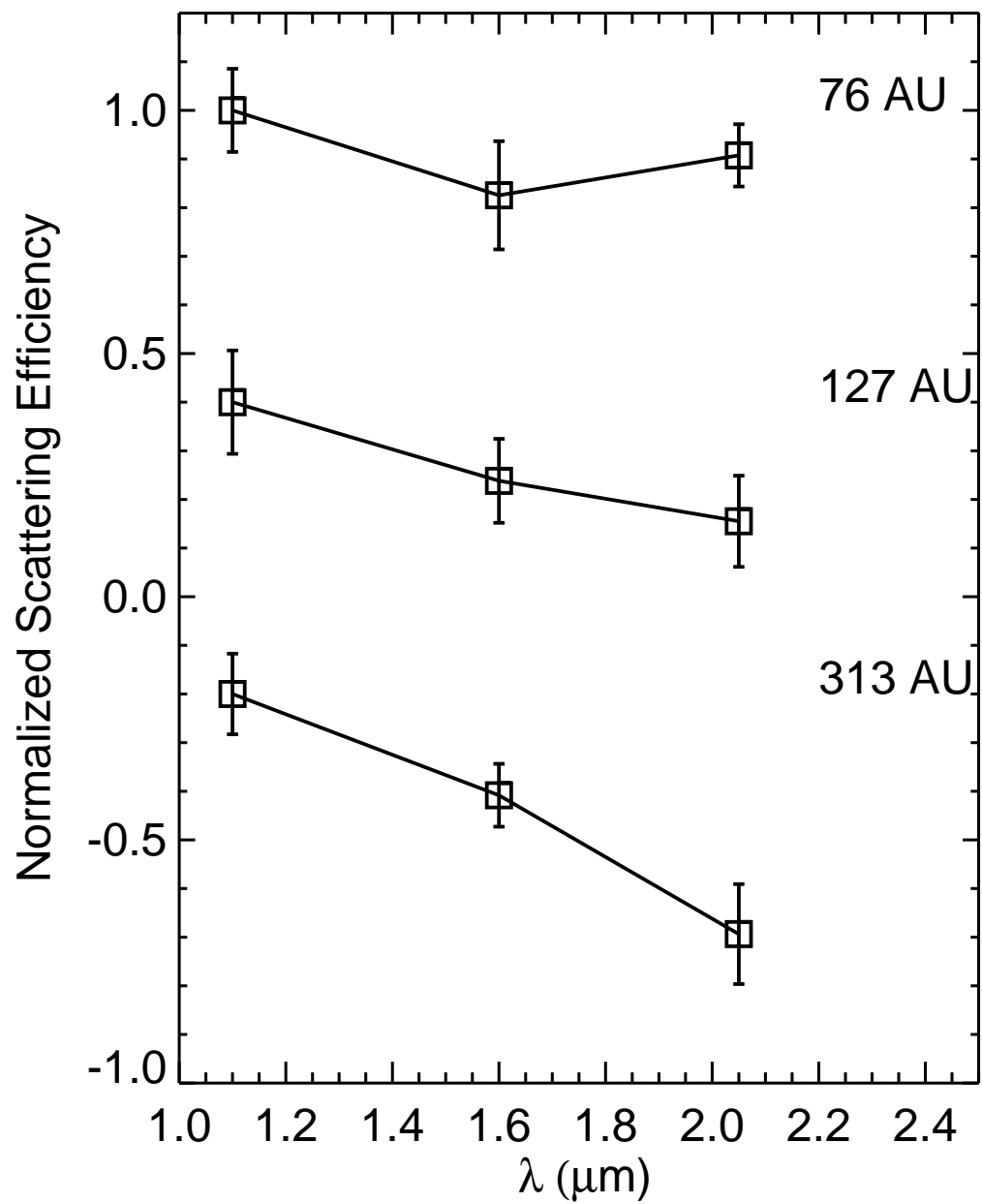}
\caption{\label{fig:color2} Same as Figure \ref{fig:color1} but for the Southwestern lobe.}
\end{figure}

We used our 3$\times$3 pixel apertures as described in \S \ref{sec:sb} to convert our surface brightness measurements into scattering efficiencies for distances from the star ranging from 0\farcs6 (67~AU) to 3\arcsec\ (337~AU).  Figures \ref{fig:color1} and \ref{fig:color2} show the scattering efficiency as a function of wavelength for 76, 127, and 313~AU, showing a representative range of colors for the disk as a function of distance.  In particular, both lobes of the disk are roughly neutral in the interior of the disk and blue at the outer edges.   The SW lobe has a red zone between 127~AU and 313~AU.

The surface brightness asymmetries appear to be the result of a color difference between the lobes.  The color change begins at $\sim$127~AU, corresponding roughly to where the warp in the Southwestern lobe becomes noticeable.    The mostly neutral colors in the near-IR favor grains in the disk that are close in size to (or larger than) the wavelength of light observed, or $\sim$1\micron.  Both silicates and water ice would be primary candidates to produce the observed efficiencies.

\section{Asymmetries and Warps Due to Interaction with the ISM}
\label{s:modeling}
The discovery of $\delta$-Velorum's high velocity interaction with the ISM in the Local Bubble \citep{gaspar08} and the peculiar morphology of the debris disk HD~61005 demonstrate the need to also consider the interaction of ISM gas with circumstellar material \citep[][ hereafter H07]{hines07}.  A general picture of how the ISM may affect circumstellar disks is nicely summarized in \citet[][hereafter AC97]{artymowicz97} as well as in H07 and references therein.  We argue that the simplest explanation for the color and brightness asymmetries, as well as the perturbed morphology of the outer disk, is the interaction of HD 32297 with a cloud of ISM material as first suggested by K05.  

The behavior of HD~32297's surface brightness profiles are similar in behavior to the disks of AU Microscopii and $\beta$-Pictoris, both of which show breaks in their radial brightness profiles \citep{heap00,golimowski06,liu04,kalas04,metchev05,graham07,fitzgerald07b}. These breaks are interpreted as arising from the ``birth-ring'' of large planetesimals that feed both the inner dust population through Poynting-Robertson or wind drag and the outer dust population through radiation pressure and mutual collisions \citep{strubbe06}.  The outer disk is then composed of increasingly weakly bound dust grains, or grains that are unbound and leaving the system (so-called $\beta$ meteoroids).

  There are two components to an ISM cloud that dust grains both bound and unbound in a disk will interact with.  First is the hydrogen gas which makes up the bulk of the mass in a cloud, and thus is the dominant perturbation on the dynamics of the grains through supersonic ballistic gas drag.  The second component is the small ISM grains in the cloud which will primarily affect the composition and grain size distribution of the disk grains through collisions with dust in the disk.  ISM grains are repelled by radiation pressure of the central star at some radius, so this effect is primarily in the outer regions of a debris disk (AC97).  Both of these components can have an observable affect on the morphology of the disk.  In this Section we develop for the first time a simple model for a disk interacting with the gaseous part of the ISM.  We apply it to HD~32297 and two other disks that have peculiar morphologies, HD~15115 and HD~61005.

\subsection{The Interaction of Dust Grains with ISM Gas}

 We consider the situation where a collisional cascade has resulted in a steady flux of unbound $\beta$ meteoroids traveling at escape speed from the system and encountering the ISM gas.  Since the relative velocity of the dust grains is on the order or greater than the sound speed of the gas up to a gas temperature of $\sim$8000~K (the maximum temperature for a neutral cloud), the interaction of a single dust grain with the HI gas in the neutral ISM is akin to supersonic gas drag felt by meteorites entering the Earth's atmosphere with a drag coefficient $D_g$ \citep{pecina83,hoppe37}.  In this model, the acceleration ($\mathbf{\vec{a}}$) a single dust grain feels from the gravity of the central star, the radiation pressure from the central star, and the drag from the ISM gas at a distance $r$ is \citep[e.g.][]{nakagawa86} 
\begin{equation}
\label{eq:acc}
\mathbf{\vec{a}}=-\frac{GM_\star(1-\beta)}{r^2}\mathbf{\hat{r}}-D_g \left(\mathbf{\vec{v}_{\mathrm{dust}}}-\mathbf{\vec{v}_{\mathrm{gas}}}\right)^2
\end{equation}
where the drag coefficient is \citep{hoppe37,pecina83}

\begin{equation}
\label{eq:dg}
D_g = \frac{A\Gamma\rho_{gas}}{\rho_{\mathrm{dust}}^{2/3}m_{\mathrm{dust}}^{1/3}}.
\end{equation}
Here, $A$ is a shape parameter and $\Gamma$ is the drag coefficient for a supersonic spherical particle ($A\Gamma\sim$1), $\rho_{gas}$ and $\rho_{dust}$ are the densities of the ISM gas and the disk dust grains respectively, $\mathbf{\vec{v}_{\mathrm{dust}}}$ and $\mathbf{\vec{v}_{\mathrm{gas}}}$ are the velocities of a dust grain and the ISM flow respectively.  We define $\beta$ as the ratio of the force of radiation pressure to the gravitational force exerted by the star. 

An order of magnitude estimate for the radial distance at which the dust grains are significantly perturbed from their original trajectories can be obtained by setting $\mathbf{\vec{a}}$=0 and assuming weakly bound grains, i.e. $\beta\sim$0.5, in Equation \ref{eq:acc}.

\begin{equation}
\label{eq:deflect}
R_{\mathrm{deflect}} \simeq \left(\frac{GM_\star}{2 D_g}\right)^{\frac{1}{2}} v_{\mathrm{rel}}^{-1}
\end{equation}

where $v_{\mathrm{rel}}$ is the relative velocity between the disk and the ISM.  The assumption of $\beta\sim0.5$ implies a particular grain mass and density for a given dust grain composition.   $\beta$ for a perfectly absorbing spherical dust grain with radius $r_{\mathrm{dust}}$ is \citep{burns79}

\begin{equation}
\label{eq:beta}
\beta = 0.574 \left(\frac{L_\star}{M_\star}\frac{M_\odot}{L_\odot}\right)\left(\frac{1 \mathrm{g/cm^3}}{\rho_{\mathrm{dust}}}\right)\left(\frac{1 \micron}{r_{\mathrm{dust}}}\right).
\end{equation}

We can investigate $R_{\mathrm{deflect}}$ as a function of $\rho_{\mathrm{gas}}$, $v_{\mathrm{rel}}$, and $L_\star$ for a star moving through a typical ISM cloud.  If we combine Equations \ref{eq:deflect} and \ref{eq:beta} for $\beta$=0.5 we remove the dependence on both dust grain size and density and create a scaling law for $R_{\mathrm{deflect}}$:

\begin{equation}
\label{eq:deflect2}
R_{\mathrm{deflect}} = 286 \left(\frac{1.67\times10^{-22} \mathrm{g~cm^{-3}}}{\rho_{\mathrm{gas}}}\right)^{\frac{1}{2}}\left(\frac{20 km s^{-1}}{v_{\mathrm{rel}}}\right)\left(\frac{L_\star}{L_\odot}\right)^{\frac{1}{2}} \mathrm{AU}
\end{equation}

To synthesize model debris disk images to compare to our data, we integrated Equation \ref{eq:acc} for various dust and cloud parameters using a modified N-body integrator.  We released thousands of particles with initial velocities determined by the birth ring orbital parameters: with an initial semi-major axis, uniformly distributed inclinations from 0$^\circ$ to 5$^\circ$, uniformly distributed eccentricities $<$0.1, uniformly distributed longitudes of ascending node, uniformly distributed longitudes of pericenter, and true longitudes uniformly distributed between 0 and 2$\pi$.  We assumed the dust grains have $\beta$=0.51, that is, they are about the blow-out size for the star.  A more thorough model could self-consistently calculate $\beta$ for a given composition and grain size.  This approach is unjustified for most debris disks as there are currently few compositional constraints. 

We calculated each particle's trajectory with a Bulirsh-Stoer integrator \citep{numrec} with an initial timestep of 5 years, for a total of 5000~yr.  We recorded particle positions every 5 years and accumulated them in a two dimensional histogram of projected particle density

We constructed a final disk image by calculating the surface brightnesses of the columns of particles in each pixel.   We then Gaussian smoothed the resulting image to model convolution with the PSF of a telescope.  For example, a simulation of a NICMOS image of HD~32297 at F160W has a pixel size of 75 mas (corresponding to 8.5~AU at HD~32297's distance) and the image uses a Gaussian smoothing kernel with a FWHM of $\sim$137 mas.  This method of using histograms to represent the steady-state distribution of dust clouds have been used by several other authors \citep{wilner02, moromartin03,moran04,stark08}.

\subsubsection{An ISM Gas Interaction Model for HD~32297}
\label{s:modeling:hd32}
Modeling the structure of the HD~32297 disk ab initio would require knowing the mass, $T_{eff}$,and luminosity of the star, the relative motion of the cloud compared to the disk, the composition of the disk grains, and the density of the gas cloud, all of which are poorly constrained or unknown.  Therefore, we have simply chosen initial conditions to qualitatively match the observed morphology.  With this approach we can 1) demonstrate that this scenario can plausibly explain many features within the disk using reasonable initial conditions, and 2) back out some weak constraints on the disk grains and cloud parameters that appear to fit the data.

Before we experimented with the dynamical models, we re-examined the stellar parameters of HD~32297.  Literature values for the spectral type to HD~32297 range from A0 \citep{hip} to A5 \citep[][and references therein]{fitzgerald07}.  We conducted a least-squares fit of Kurucz models to HD~32297's BVJHK$_s$ photometry to determine both $T_{eff}$ and luminosity.  The model that best fits the photometry has T$_{eff}$=7750 K and luminosity = 5.7 L$_\odot$.  

However, these values place HD 32297 below the
ZAMS on the tracks of both \cite{siess00} and \cite{dantona87}, i.e. in
an unphysical regime. The 1-$\sigma$\ uncertainty on the distance to the star in the new Hipparcos
reduction of \cite{vanl} is 12 pc. Keeping the star at 7750~K
and moving it farther away by 1$\sigma$, i.e. from 112 pc to
133 pc, would increase the inferred luminosity to 8.1 L$_\odot$; this is the luminosity on the ZAMS for a star of 1.7 M$_{\odot}$ on the Siess tracks. We take these values to be the most likely stellar luminosity and mass.

One caveat to the above analysis is the possibility of extinction from the intervening ISM or circumstellar material.  If the intrinsic T$_{eff}$ of the star is closer to $\sim$10000 K, as would be
appropriate for its earliest literature spectral type of A0, then considerable
extinction would be necessary to fit the photometry. An ISM-like A$_{\mathrm{v}}$=0.72 also provides a good fit to
the observed BVJHKs photometry for T$_{eff}$=10000K and log~g=4.5 Kurucz models. In
this case the luminosity at the nominal distance is 13.8 L$_{\odot}$, which is also
well below the ZAMS at that T$_{eff}$.  At T$_{eff}$=10000K, the ZAMS has luminosity
$\sim$30 L$_{\odot}$ for a M$\sim$2.3 M$_{\odot}$.  That would require moving the star out
to 165 AU, or 4.4$\sigma$ on its Hipparcos distance.   Given the low ISM Na column density toward the star \citep{redfield07}, \citet{fitzgerald07} dismissed
the possibility of high A$_{\mathrm{v}}$, but other possible A$_{\mathrm{v}}$, T$_{eff}$
combinations could result in an allowed luminosity. 

Alternatively, the star
could experience gray extinction from circumstellar material to explain the discrepancy between the inferred luminosity and the ZAMS luminosity.  This, however, would imply an optically thick circumstellar disk comprised of fairly large dust grains. This situation seems unlikely given the low L$_{IR}$/L$_\star$ measured which is more in line with optically thin dust disks \citep{fitzgerald07}.  Also, the inferred amount of circumstellar gas is small, arguing for a late stage of evolution for the disk where relatively little primordial gas and dust remain \citep{redfield07}.

In addition to the stellar parameters, we must consider the relative velocity vector between the disk and the ISM flow, $\beta$, and $D_g$ to fully determine the initial conditions for our simulations.  The relative velocity vector can be weakly constrained by the local dispersion of ISM cloud velocities and HD~32297's proper motion.  The other parameters are more difficult to constrain.
 
In order to determine a $v_{\mathrm{rel}}$ consistent with both the observed proper motion and radial velocity of HD~32297 as well as the morphology of the warped disk, one must make some assumptions about the impacting cloud's velocity vector relative to that of the star.  From observations of the circumstellar and interstellar Na lines in HD~32297's spectrum the radial velocity of HD~32297 is roughly 20 km/s while the ISM's radial velocity is $\sim$ 24 km/s \citep{redfield07}.  HD~32297's proper motion is primarily to the South at a speed of 15 km/s (assuming a distance closer to 133~pc).  If the cloud only moves in the radial direction, the magnitude of the relative velocity between the cloud and HD~32297 would be 16 km/s.  The cloud may have additional proper motion relative to HD~32297.  If we assume that the cloud's velocity vector is parallel to HD~32297's motion on the sky that would produce a minimum relative velocity, while if it was moving anti-parallel it would produce the maximum relative velocity.  Based on the dispersion in velocity of nearby ISM clouds, we assume the magnitude of the cloud's velocity relative to that of HD~32297 is $\sim$14~km s$^{-1}$ \citep{redfield08}, the minimum and maximum relative velocities between the cloud and HD~32297 would be 4~km s$^{-1}$ and 29~km s$^{-1}$.  

We varied the relative velocity vector and $D_g$ until we found qualitatively similar morphologies between our models and HD~32297.
The best morphological match came from a relative velocity vector with a magnitude of 15~km/s at an inclination of 55$^\circ$ with respect to the y-z plane in coordinates where the x-axis is aligned with the disk midplane and the z axis is perpendicular to the midplane.  This represents parameters in the middle of our estimates above.  

For this model, $D_g$=1.2$\times$10$^{-18}$ cm$^{-1}$. Given the above parameters, one would expect $R_{\mathrm{deflect}}\sim$438~AU, about where the southwestern lobe of the disk shows a warp of $\sim$7$^\circ$. 

\begin{figure}
\plottwo{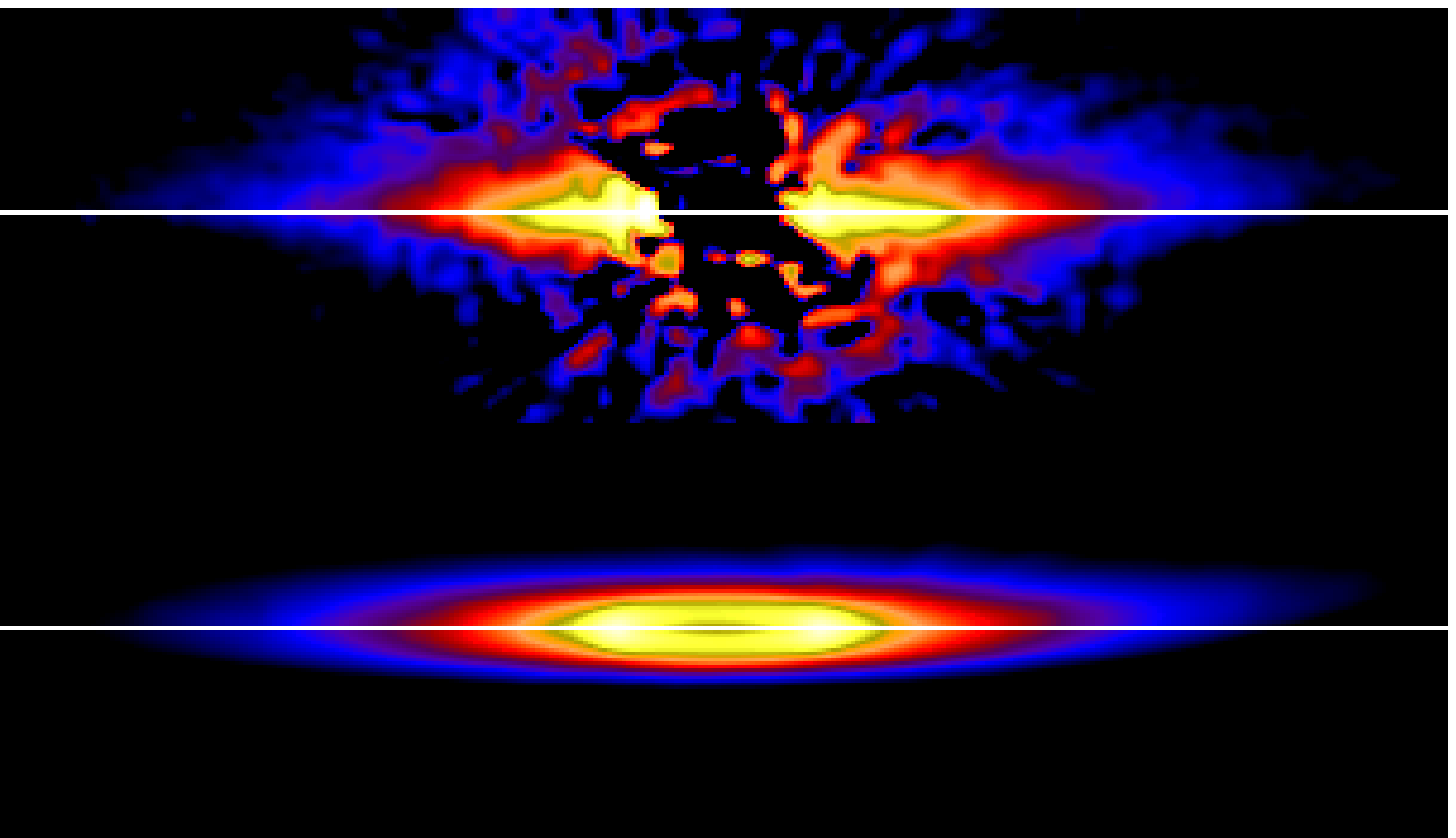}{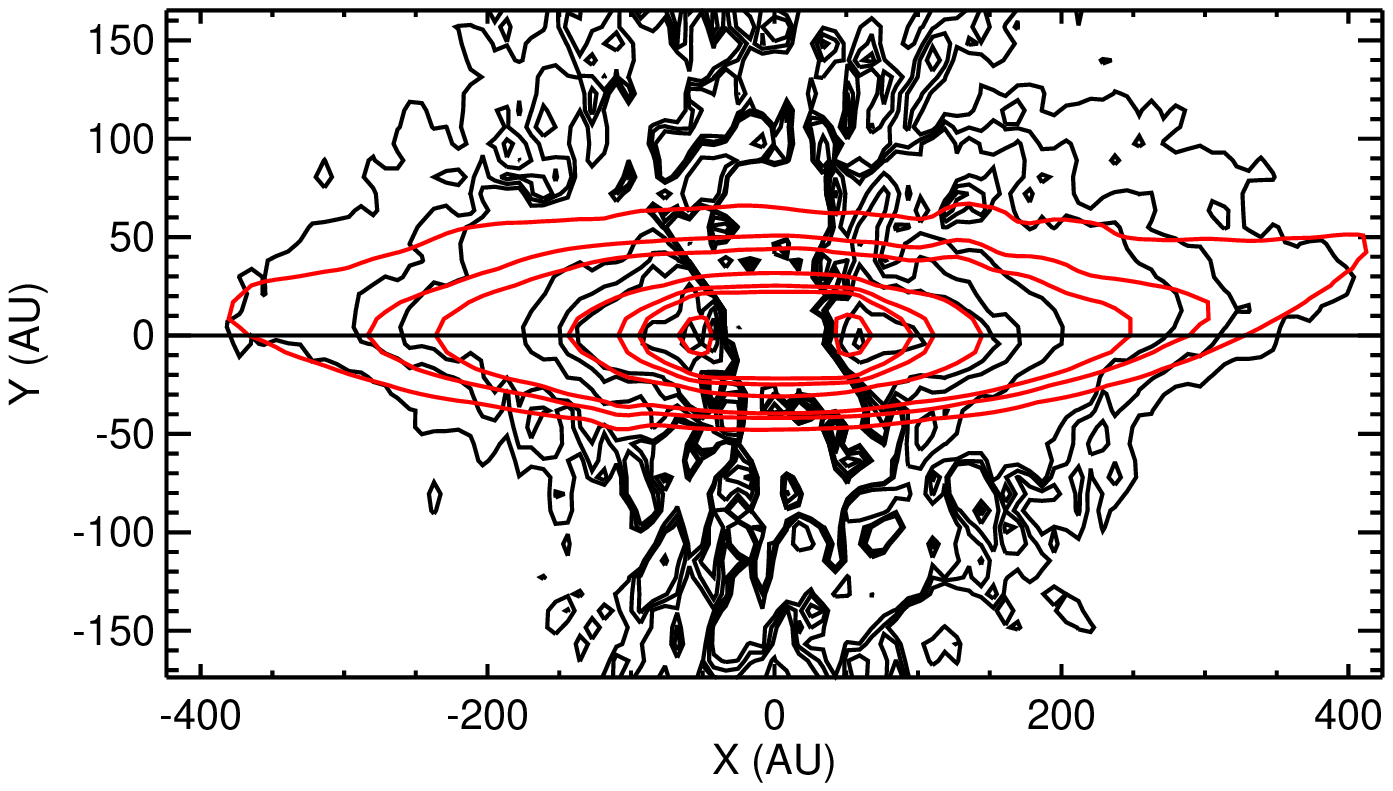}
\caption{\label{fig:comp} (left) Comparison between an F160W image of HD~32297's disk and a model of a disk interacting with a cloud of ISM gas with the parameters mentioned in Section \ref{s:modeling:hd32}. (right) A comparison of surface brightness contours between HD~32297 (black lines) and our model (red lines).  Our model has been arbitrarily scaled to match the maximum observed surface brightness.}
\end{figure}

  We show a comparison between the F160W image of HD~32297 and our synthetic image in Figure \ref{fig:comp}, arbitrarily scaled to similar maximum brightnesses and using the same stretch.  We also show a comparison between two contour maps of the surface brightness, using the same scaling.  The model and the data agree qualitatively.  

\subsubsection{An ISM Gas Interaction Model for HD~15115}
\label{s:modeling:hd15}

HD~15115 is an F2 star at 45~pc that has a disk with a highly asymmetric edge-on morphology; the Eastern lobe of the disk appears truncated relative to the Western lobe \citep{kalas07,debes08}.  The proper motion vector of HD~15115 is southeast, with its eastern component almost twice as large as its southern component.  In this case, we do not have a measure of the ISM cloud's radial velocity, so we strictly consider the relative transverse motion in the sky between the cloud and HD~15115. We can convert this transverse motion to determine the relative velocity constraints between the disk and the hypothesized ISM.  The disk midplane is oriented with a PA of 279$^\circ$.  The transverse velocity vector of HD~15115 on the sky is at roughly the same PA as the disk.  If we again assume a cloud with velocity of of 14~km/s on the sky, the minimum and maximum possible relative transverse velocity is between 7 and 35 km/s if the cloud's velocity vector is parallel or anti-parallel to HD~15115's proper motion.  The morphology of the disk is consistent with the midplane plowing headlong into an ISM flow.  

We assumed that HD~15115 had a stellar mass of 1.5 M$_\odot$ and that grains were launched from 35~AU, the rough inner radius of HD~15115's disk as inferred from its IR excess \citep{zuckerman04,williams06}.  Experimenting with different models showed that $D_g$=1.3$\times$10$^{-18}$ cm$^{-1}$ and a relative velocity vector with equal components along the midplane and along the observer's line of sight with a magnitude of 30 km/s best matched the observed data.  

\begin{figure}
\plotone{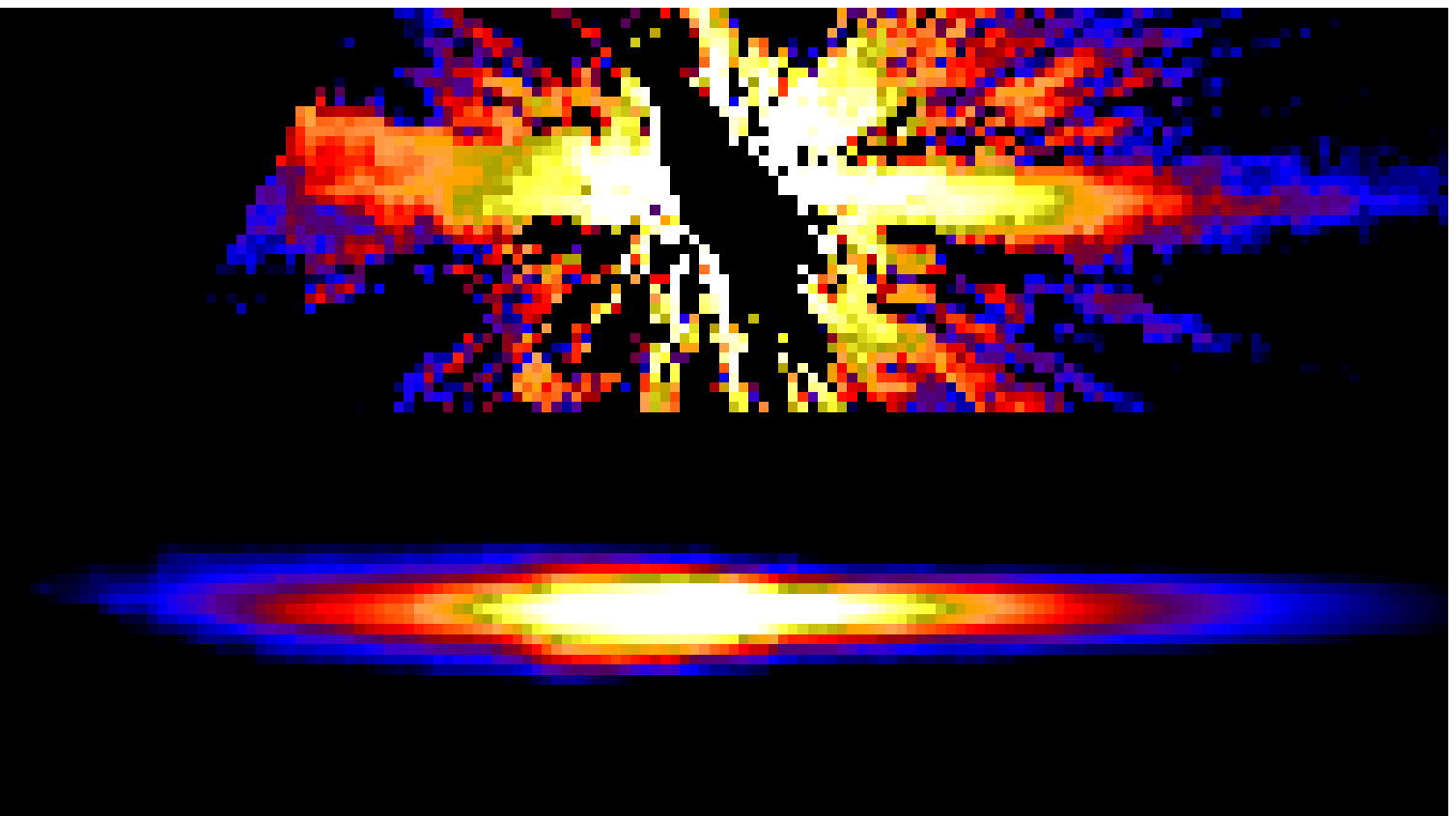}
\caption{\label{fig:hd15} Comparison between an F110W image of HD~15117's disk from \citet{debeshd} and a model of a disk interacting with a cloud of ISM gas with the parameters mentioned in Section \ref{s:modeling:hd15}.}
\end{figure}

Figure \ref{fig:hd15} shows a comparison of the F110W image of HD~15115 with our disk model on the same size scale.  The eastern lobe (left side of Figure \ref{fig:hd15}) is contaminated by a strong residual that points roughly northeast in the Figure \citep[see][]{debes08}.  Otherwise, the figure shows that the morphology of the model matches the observation of HD~15115 showing a significant truncation of one lobe of the disk.  Additionally the figure shows that our models predict that the FWHM of the disk midplane should be wider on the truncated side.  This aspect of the disk morphology is not significantly probed by current observations \citep{debes08}.  Another proposed scenario for this morphology is the close fly-by of HD~15115 by another star \citep{kalas07}.  However, no definitive candidates for a fly-by have currently been detected \citep{debes08}.

\subsubsection{An ISM Gas Interaction Model for HD~61005}
\label{s:modeling:hd61}
The disk observed around HD~61005, the ``Moth'', shows a peculiar disk morphology reminiscent of material being pushed up out of the disk midplane (H07).  This shape might indicate disk interaction with the gas component of the ISM.  The star itself is a late G8 dwarf at a distance of 34~pc, and H07 found that the surface brightness of the dust is consistent with dust grains roughly the size of the blowout radius, 0.3~\micron.  H07 also determined from HD~61005's disk SED that the grains originate from radii $\sim$10~AU from the star.   

The proper motion of HD~61005 is -56.09 mas in RA, and 74.53 mas in declination \citep{vanl}.  If we rotate these velocities to a frame where the x-axis is parallel to the presumed major axis of the disk and convert to relative velocities under the same assumptions we used above as for HD~15115, the relative velocities will range in magnitude from 2~km/s to 30~km/s, with a vector primarily along the x-z plane at an inclination of 68$^\circ$ relative to the x-axis.  If icy 0.3\micron\ grains are the population we observe around HD~61005, we can derive $D_g$ under the assumption of an ISM gas density N(HI)=100 cm$^{-3}$.  These two assumptions result in $D_g=3.3\times10^{-18}$ cm$^{-1}$

\begin{figure}
\plotone{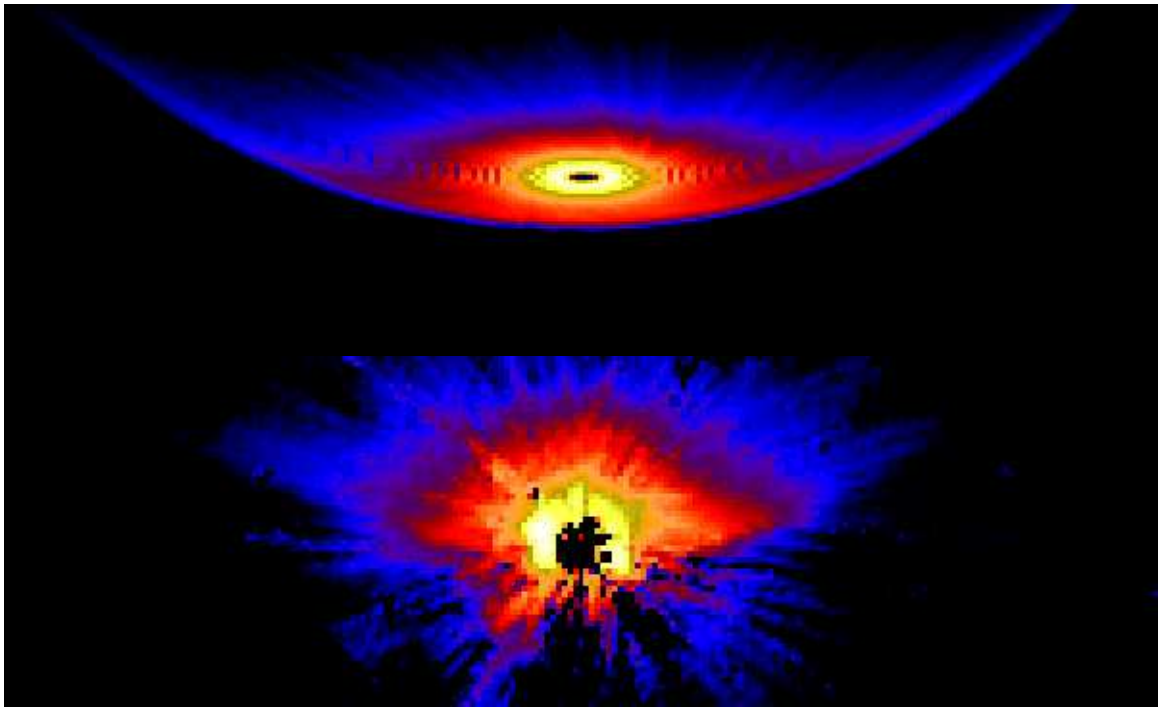}
\caption{\label{fig:hd61} Comparison between an F110W image of HD~61005's disk (kindly provided by G. Schneider) and a model of a disk interacting with a cloud of ISM gas with the parameters mentioned in Section \ref{s:modeling:hd61}.}
\end{figure}

We experimented with various relative velocity vectors until we found a qualitative match to the data.  A velocity vector that is nearly perpendicular to the disk midplane with a magnitude of 25~km/s matches the data best, along with an inclination between the observer and the disk of 20$^\circ$.
Figure \ref{fig:hd61} shows a comparison of the model and the data, at the same size scale and stretch.  We also see good qualitative agreement with the morphology of this disk as well.


\section{Discussion}

In this Section we discuss some details that our ISM interaction model presently lacks.

\subsection{ISM Dust Grains}

For a disk interacting with an atomic ISM cloud with small grains, AC97 calculated an ``avoidance radius'', $r_{av}$, where the radiation pressure of the central star was strong enough to scatter small grains away from a circumstellar disk.  The distance $r_{av}$ depends on $\beta$, $v_\infty$, the initial velocity of a grain at large distances, and the stellar mass $M_\star$.
\begin{equation}
r_{\mathrm{av}} = \frac{2(\beta - 1) GM_\star}{v^2_\infty}.
\end{equation}

Incoming ISM dust grains  pile up at this radius and scatter away from the star, leaving a paraboloidal region with minimum radius $r_{\mathrm{av}}$ that is free from sandblasting from small ISM grains. 
This boundary marks the separation between where grains from the ISM could have a significant effect and where the ISM dust does not impact the evolution of dust in a circumstellar disk. 

 The color asymmetries observed for HD 32297 could potentially be explained by this ISM grain sandblasting.  With some assumptions about the ISM grains as well as the cloud that may be impacting HD~32297's disk, we can estimate what $r_{av}$ is for HD~32297.  AC97 calculated expected $\beta$'s for ISM dust for an A5 star such as HD~32297.  They calculated a maximum $\beta$ for 
$\beta$-Pictoris assuming silicate+carbonaceous grains of 19.2. HD~32297 has a
similar luminosity to $\beta$-Pictoris, so  $r_{av}$ for HD~32297 would then be on order of between 133~AU and 1061~AU.  The relative velocity of 15~km/s that we used for our gas interaction model of HD~32297 would correspond to an $r_{\mathrm{av}}$=245~AU.  This is slightly larger than expected from the observations, and could be accounted for by a combination of the gas interaction and sandblasting by ISM grains.  

AC97 calculated how ISM grain sandblasting would affect dust erosion rates in a disk.  They found that there should be larger dust removal rates in the part of the disk directly impacting the ISM, which in turn causes a large asymmetry in dust production/erosion between regions that suffer direct sandblasting and those that avoid it.  Additionally, there is an asymmetry between dust grains that are moving anti-parallel with the ISM flow and those that move away from the ISM flow, especially for grains in orbit around the star.  Detailed modeling of the interaction with self-consistent grains of a particular composition and between the dusty ISM grains as well as the HI gas contained in the ISM cloud is left for a future paper (Debes, Kuchner, \& Weinberger, 2009, in prep).

\subsection{Other Uncertainties in the ISM Interaction model}
\label{sec:caveats}

While the interaction of dust with the surrounding ISM provides an attractive explanation for the above disks, our models contain several uncertainties.  For example, we have followed only a single size of dust grain in our interaction models, when clearly a size distribution of grains will be present in the disk.  Furthermore, we have assumed a steady bulk flow to the ISM, even though one would expect turbulence on moderate size scales which could create further substructure.  

While we found specific values of $D_g$, $\beta$, and $v_{rel}$ that fit the observed morphology, any of these can be varied to balance out a change in the other; we cannot derive unique fits to the data without further constraints on composition and $v_{rel}$.  Choosing a specific composition and grain type, such as spherical or fluffy particles would fix $\beta$ and constrain $D_g$.  Modeling the morphology and surface brightness of a disk at multiple wavelengths may provide further constraint on a particular gas interaction model.  In some cases this effect should help to further constrain dust grain particles, since the exact morphology of an interaction depends on the dust properties.

Physical processes, such as the presence of a stellar wind, can work to frustrate the interaction between the ISM and the circumstellar disk.  For young solar-type stars, such as HD~61005, this may be important, while this may be less so for earlier type stars that have less magnetic activity.  Regardless, models of interactions between a dense ISM cloud and the solar wind can guide us to the order of magnitude of this effect.  \citet{yeghikyan03} modeled the interaction between an ISM cloud with N(HI)=100 cm$^{-3}$ and a solar-type wind.  They found that when the sun was placed in such a dense cloud the heliopause shrunk to $\sim$1~AU, well interior to the circumstellar region we observe for HD~61005.  However, younger stars have stronger winds, which shield their debris disks out to hundreds of AU.

Finally, there are large uncertainties in the velocities and densities of the modeled ISM clouds.  Background stars near each of the above stars could provide a possible probe of the clouds and constrain their density/radial velocity by providing lines of sight spatially coincident with the debris disks.  Measurement of the transverse motions of the clouds would require a way of detecting the clouds directly at high spatial resolution, a difficult endeavor.
  
\section{Conclusions}
\label{s:conc}

We have detected the HD~32297 debris disk at two additional wavelengths, bringing to four the total number of wavelengths in which this disk has been observed in scattered light.  Both the morphology and the surface brightness of the disk as a function of wavelength hold useful clues to the nature of the dust in orbit around the star.  The addition of more resolved near-IR scattering images of the HD~32297 disk provides measures of the morphology, surface brightness profiles, and 
scattering efficiency of the dust at three separate wavelengths.  

We find that the morphology and asymmetries of the HD~32297 disk can qualitatively be explained by the interaction of the disk's dust with a moderately dense ISM cloud.  We developed a numerical model for the ineraction of a debris disk with ISM gas, and applied it to this system.  We have also applied our ISM interaction model to two other disks, HD~15115 and HD~61005 and demonstrated that some warps and asymmetries in the outer regions of these disks can plausibly be explained by this mechanism.

Future coronagraphic observations of HD~32297' debris disk, especially with STIS, will precisely measure the scattering efficiency of the inner disk in visible wavelengths. More work needs to be done to understand precisely the effect of a debris disk's environment on how it appears.


\acknowledgements
The authors would like to thank Dean Hines and Glenn Schneider for kindly providing an image of HD~61005 and useful discussions on both HD~61005 and HD~32297.  We would also like to thank Carey Lisse for pointing out that stellar winds can affect dust/gas interactions and Chris Stark for a careful reading of the manuscript.  This research is based on observations with the NASA/ESA Hubble Space Telescope which is operated by the AURA, under NASA contract NAS 5-26555. These observations are associated with programs GO-\#10177, \#10527, and \#10857.  Support for program \#10857 was provided by NASA through a grant from STScI.  This research was supported by an appointment of JD to the NASA Postdoctoral Program at the Goddard Space Flight Center, administered by Oak Ridge Associated Universities through a contract with NASA.  AJW and MJK also acknowledge support from the NASA Astrobiology Institute. 

\bibliography{scibib}
\bibliographystyle{apj}

\begin{deluxetable}{cccc}
\tablecolumns{4}
\tablewidth{0pc}
\tablecaption{\label{tab:pl} Power-law indices to the Surface Brightness Profiles}
\tablehead{
\colhead{Filter} & \colhead{R$_{\mathrm{break}}$ (AU)} & \colhead{Inner Index ($\alpha$)} & \colhead{Outer Index ($\beta$)} 
}
\startdata
\multicolumn{4}{c}{Southwest Lobe} \\
F110W & 104$\pm$6 & -1.4$\pm$0.2 & -3.33$\pm$0.03 \\
F160W & 111$\pm$3 & -1.0$\pm$0.2 & -3.57$\pm$0.03 \\
F205W & 115$\pm$4 & -0.4$\pm$0.2 & -3.9$\pm$0.1  \\
\multicolumn{4}{c}{Northeast Lobe} \\
F110W & 80$\pm$10 & -1$\pm$0.7 & -3.00$\pm$0.05 \\
F160W & 90$\pm$10 & -1.8$\pm$0.3 & -3.33$\pm$0.05 \\
F205W & 97$\pm$6 & -1.6$\pm$0.4 & -3.8$\pm$0.06 \\
\enddata
\end{deluxetable}

\end{document}